\documentclass[pre,twocolumn,superscriptaddress,floatfix,showpacs,showkeys]{revtex4}

\usepackage[latin2]{inputenc}
\usepackage{amsmath}
\usepackage{amssymb}
\usepackage{epsfig}

\newcommand{\Fcrit}{F_\text{crit}}
\newcommand{\Fcont}{F_\text{cont}}

\begin{document}

\title{Unjamming of Granular Packings due to Local Perturbations: Stability
and Decay of Displacements}

\date{\today}

\author{M.\ Reza Shaebani}
\affiliation{Department of Theoretical Physics, Budapest Univ.\ of 
 Techn.\ and Econ., H-1111 Budapest, Hungary}
 \affiliation{Institute for Advanced Studies in Basic Sciences, Zanjan 45195-1159, Iran}
\author{Tam\'as Unger}
\affiliation{Department of Theoretical Physics, Budapest Univ.\ of 
 Techn.\ and Econ., H-1111 Budapest, Hungary}
\affiliation{Solid State Research Group of the HAS, Budapest Univ.\ of 
 Techn.\ and Econ.}
\author{J\'anos Kert\'esz}
\affiliation{Department of Theoretical Physics, Budapest Univ.\ of 
 Techn.\ and Econ., H-1111 Budapest, Hungary}
\affiliation{Solid State Research Group of the HAS, Budapest Univ.\ of 
 Techn.\ and Econ.}

\begin{abstract}
We study the mechanical response generated by local deformations in jammed
packings of rigid disks. Based on discrete element simulations we determine
the critical force of the local perturbation that is needed to break the
mechanical equilibrium and examine the generated displacement
field. Displacements decay as a power law of the distance from the
perturbation point. The decay exponent and the critical force exhibit
nontrivial dependence on the friction: Both quantities are nonmonotonic and
have a sharp maximum at the friction coefficient 0.1. We find that the
mechanical response properties are closely related to the problem of
force-indeterminacy where similar nonmonotonic behavior was observed
previously. We establish direct connection between the critical force and
the ensemble of static force networks.
\end{abstract} 



\maketitle

Granular materials constitute an ideal field for studying the physics of
the jamming transition \cite{Liu01,Majmudar07}.  One of the most exciting
challenges of research is to provide better understanding of the onset of
yielding in granular media, an example of unjamming.  When the external
load on a static assembly of grains is changed at a certain point the load
may become incompatible with the inner structure of the packing and the
solid state looses its stability. How exactly this happens on grain-scale
and what are the key features of the transition between statics and flow
are intriguing and unresolved problems
\cite{JNRoux00,Combe00,Kolb06,McNamara05}.  Moreover, it is of essential
importance in many applications to be able to predict, initiate or prevent
such transitions.

In this Letter we study yielding induced by local perturbations and focus
on two response properties. First, we investigate the resistance of the
packing that is exerted against a local deformation. Second, based on the
generated displacement field, we study how far the effect of the
perturbation penetrates into the packing. In recent years local
perturbations were extensively used to study the nature of stress
transmission in granular media \cite{Geng01,Goldenberg05,Ostojic06}. These
perturbations were typically weak in the sense that they did not break the
stability: The original contact network can maintain the perturbation,
particle displacements are reversible and occur merely due to elastic
distortions of contacts \cite{Goldenberg05}. Here we deal with rigid
(undeformable) grains with help of the simulation technique of contact
dynamics \cite{Jean99,Brendel04} therefore elastic deformations are
excluded. As opposed to the weak perturbations, in our case, plastic
deformation of the packing is initiated \cite{JNRoux00,Combe00}.  Plastic
rearrangements caused by local perturbations have been studied in recent
experiments \cite{Kolb06} where interesting power law decay of the
rearrangement field was found.

An important question we address here is what effect the
particle-particle friction $\mu$ has on the response of the packings.
This was motivated by the prediction that
response properties 
may show nonmonotonic behavior as
the function of $\mu$, i.e.\ small and large frictions lead to similar
behavior which is different from the behavior for intermediate
friction \cite{Unger05}. This idea was based on the ensemble of admissible
force configurations.

It is known that, in general case, mechanical equilibrium and Coulomb
condition do not determine contact forces uniquely
\cite{Snoijer04,Unger05,McNamara05,Ostojic06}. Consequently there is an
ensemble of force networks that satisfy these conditions in the same
contact geometry and for the same external load. It depends strongly on
friction how large the indeterminacy of individual contact forces is. E.g.\
for frictionless rigid grains the indeterminacy vanishes and the ensemble
shrinks to a single admissible force network \cite{Unger05}. It was found
for 2D packings of disks that indeterminacy of forces becomes the largest
for $\mu \approx 0.1$ \cite{Unger05}, further increase of the friction
leads again to smaller indeterminacy. It was argued in \cite{Unger05} that
the extent of force-indeterminacy must affect the stability of the packing,
because the more freedom the forces have the more possibility the packing
has to resolve changes of the load without rearrangements. This suggests
that mechanical response properties may show nonmonotonic dependence on the
coefficient of friction. The main message of the present work is that such
correlation indeed exists.

The connection between mechanical properties and force indeterminacy was
also examined from an other point of view in \cite{Snoeijer06}, where the
maximum admissible shear tress was deduced from the ensemble of
force networks for packings of frictionless soft particles.

The systems we examine are two-dimensional random packings of perfectly
rigid disks in zero gravity.  The unit of the length is set to the maximum
grain radius, radii are distributed uniformly between $0.5$ and $1$. The
number of the grains contained by the packings ranges from $500$ to
$8000$. Our numerical experiments consist of two parts. First by
compression we generate dense random packings then we probe the packings by
perturbing single contacts. We apply the contact dynamics method for both
procedures.

The compaction starts from a gas-like state with grains randomly
distributed in a square-shaped cell. Periodic boundary conditions are
applied in both directions and there are no walls in the system. Instead of
using pistons the compaction is achieved by the method of
Andersen\cite{Andersen80}: We impose constant external pressure
$p_\text{ext}$ and let the volume of the cell evolve in time. A detailed
description of the coupling between the pressure bath and the system can be
found in \cite{Andersen80}. Here we mention only the favorable feature of
this method that it gives rise to homogeneous compression and, due to
exclusion of side walls, boundary effects are avoided.

As the size of the cell decreases, grains form contacts and start building
up the inner pressure in order to avoid interpenetration. Finally a static,
jammed configuration is reached where the grains block further
compaction. In the final packing contact forces are such that they provide
mechanical equilibrium for each grain and the corresponding inner pressure
$p_\text{in}$ equals to $p_\text{ext}$.

After that we turn to the perturbation part where we choose a pair of
contacting grains and force them to move apart. 
One way of doing this would be to apply a small normal force between the
two grains and continually increase it.
It is expected that if the perturbation force
is small enough then it can be resolved by the packing without
rearrangements. If the force is increased further the yield point will be
reached where the perturbation induce sliding and/or opening of some
contacts and initiates collective rearrangements of the particles at least
in the vicinity of the perturbation point.

It is very time consuming in computer simulations to scan a region of the
perturbation force in order to find the yield point.  Therefore we apply
another method. Instead of tuning the force we control the deformation. The
idea is to bring the system immediately to the yield point by enforcing the
opening of the contact: We prescribe a small gap at the contact of
perturbation and then determine the force that is needed to fulfill this
constraint. This concept is suited very well to the contact dynamics method
where interparticle forces are handled as constraint forces, i.e.\ they are
calculated based on constraint conditions which prescribe the relative
motion of the contact surfaces \cite{Jean99,Brendel04}.  It is beneficial
to choose small gap size as we are interested in the onset of motion, how
the static structure breaks due to the perturbation. Large deformations,
e.g.\ creation of new contacts, are out of the scope of the present
study. We checked that for small gap sizes the displacement field (up to a
constant factor) and the critical perturbation force become independent of
the size of the gap. Our numerical measurements are performed in this
region, the size of the gap $g$ is set to $10^{-9}$.

With the above technique we open up the contact that is selected for the
perturbation and measure the generated displacement field and the required
perturbation force. The latter describes the strength of the system against
the local perturbation and we will refer to it as the critical force
$\Fcrit$.  Both the force and the displacement response depend strongly on
the place of the perturbation. First we check their average properties.

\begin{figure}[t]
\epsfig{figure=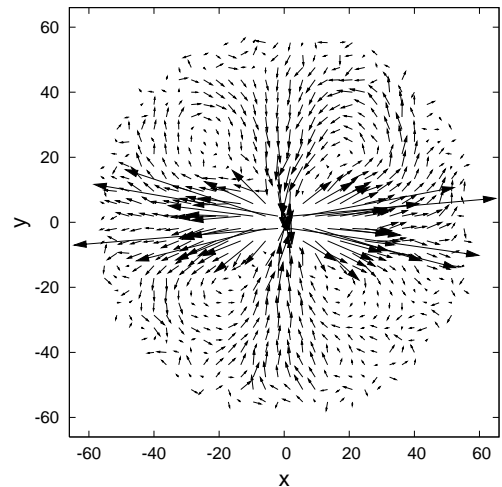,width=1.1\linewidth}
\caption{Displacement response field in the contact frame 
averaged over several thousand perturbations. The system contained $3000$
disks with friction coefficient $0.5$.} 
\label{Fig-DisplacementField}
\end{figure}

By studying the displacement field our goal is to find out how far the
rearrangements have to penetrate into the packing in order to allow the
prescribed local deformation. Is there a related length scale?  When a
single contact is perturbed the displacements of particle centers form a
disordered vector field which can be widespread or more localized depending
on the perturbed contact. In order to calculate the average displacement
field we perturb all contacts one by one always starting with the original
static packing. In each case the particle movements are recorded in the
local contact frame where the perturbed contact sits in the origin and the
$x$-axis is chosen parallel to the contact normal, i.e.\ $x$ indicates the
direction of the separation. Fig.~\ref{Fig-DisplacementField} shows a
smooth displacement field obtained by averaging over the perturbed
contacts. The apparent circular shape of the system is the consequence of
the averaging because the original square shape can have different
orientations when transformed into different contact frames.

\begin{figure}[t]
\epsfig{figure=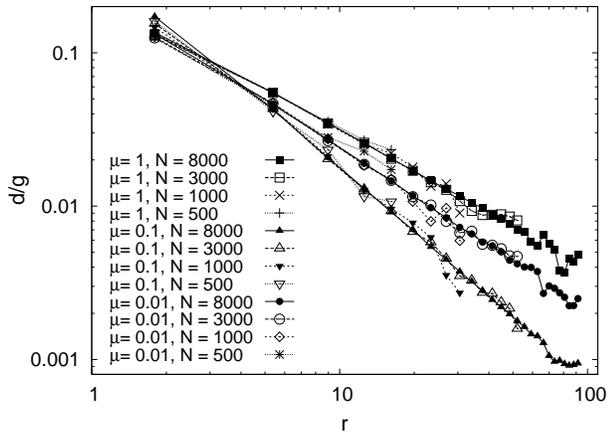,width=0.99\linewidth}
\caption{The magnitude of the displacements, $d$, 
in terms of the distance from the perturbed contact, $r$. $g$ stands for
the gap generated at the perturbed contact. Different slopes
correspond to different friction coefficients $\mu$. For each value of
friction four systems of different sizes are investigated. The total number of
particles are between $500$ and $8000$.
\label{Fig-DisplaceDistance}}
\end{figure}

The quadrupolar structure in Fig.~\ref{Fig-DisplacementField} is
interestingly very similar to the displacement fields that have been
observed in sheared systems of deformable frictionless grains
\cite{Maloney06}, where localized quadrupolar deformations appear at the
onset of plastic events.  Here, for further analysis we consider only the
magnitude of the displacement vectors and their distance $r$ from the place
of the perturbation (angle of the position is averaged out). The decay of
the the average displacement $d$ is shown in
Fig.~\ref{Fig-DisplaceDistance}.

The curves $d(r)$ show power law decay
\begin{equation}
  \label{powerlaw}
  d \propto  r^{-\alpha} \, .
\end{equation}
Thus we find that no
penetration length can be related to the rearrangements, but instead, the
penetration can be best characterized by an exponent ($\alpha$).

Variation of the system size (Fig.~\ref{Fig-DisplaceDistance}), within the
modest range allowed by the numerical tools, shows no systematic change in
the exponent. There is, however, a strong dependence on friction.  The
values we obtained for $\alpha$ are between $0.7$ and $1.4$ \footnote{For
comparison, the exponent $1$ corresponds to a two dimensional
incompressible fluid perturbed by a localized volume injection.}. Similar
power law behavior with the same range of $\alpha$ has been found
experimentally by Kolb et al.\cite{Kolb06} by moving an intruder in a
system of disks. Next we discuss the influence of
friction on the exponent and on the critical force.

\begin{figure}[t]
\epsfig{figure=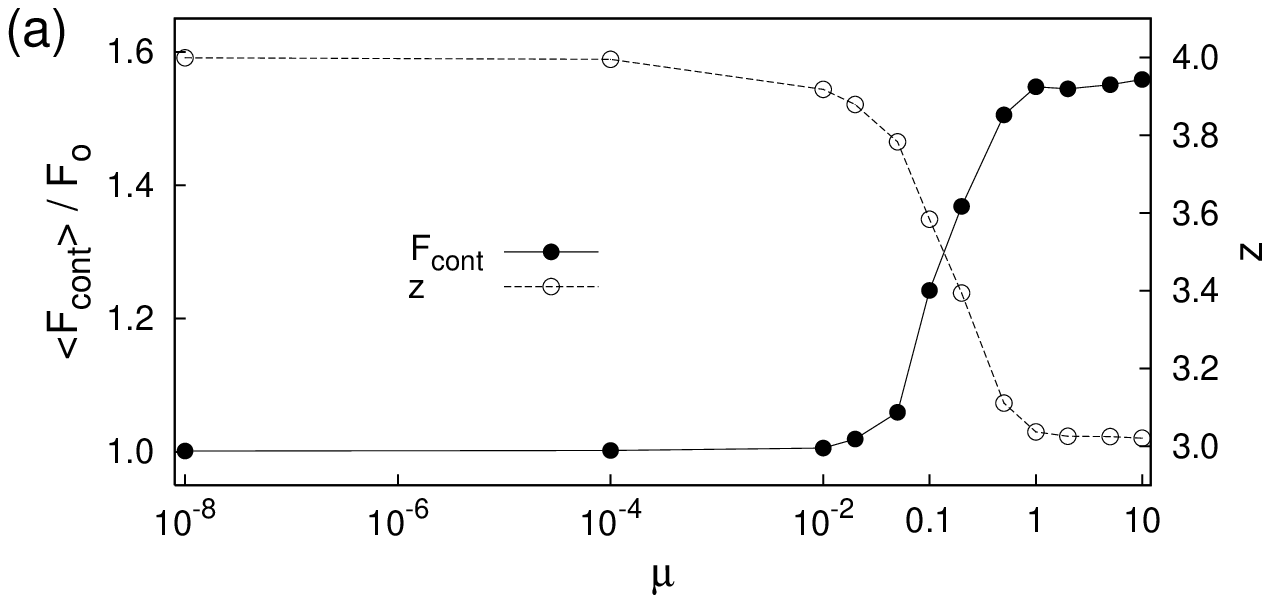,width=0.99\linewidth}
\epsfig{figure=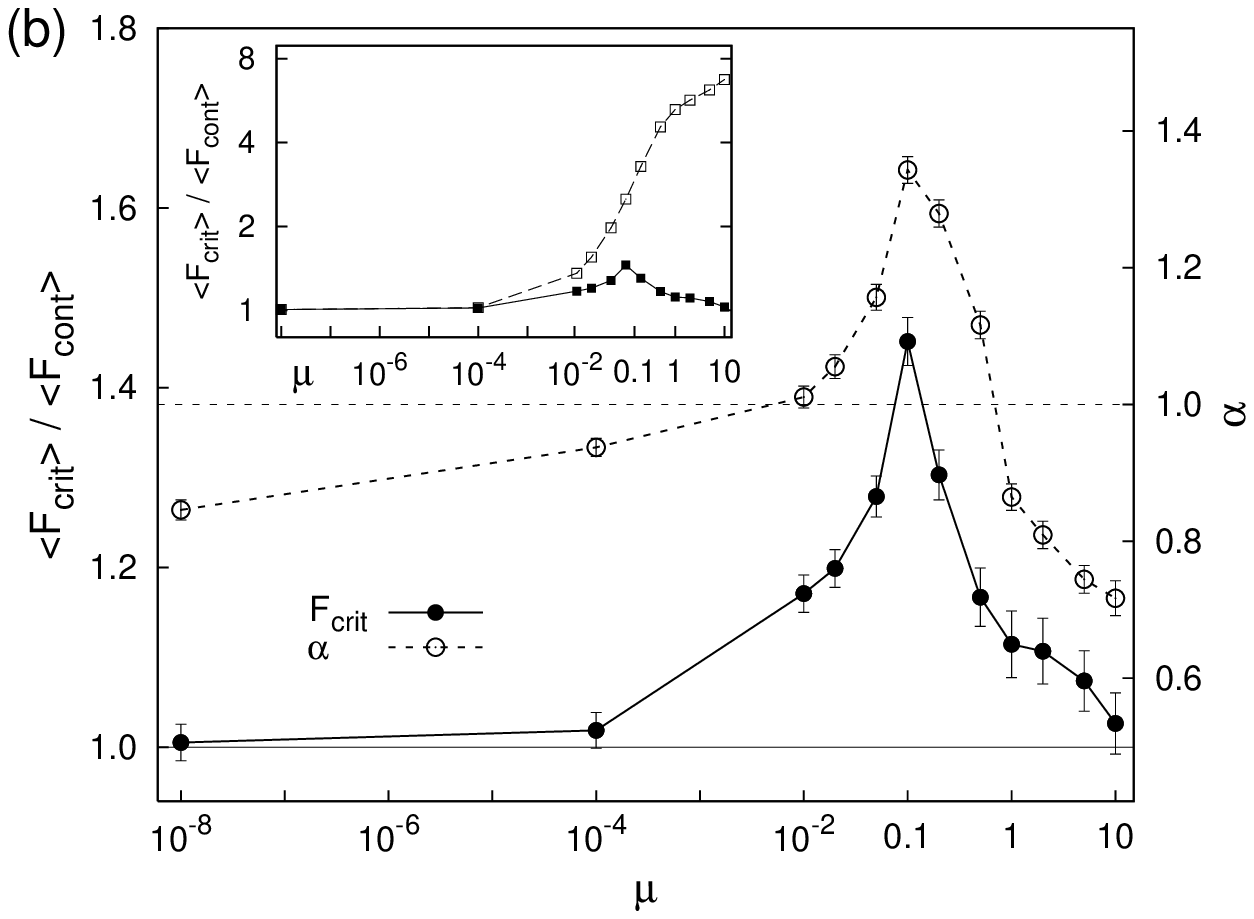,width=0.99\linewidth}
\caption{(a) Influence of friction on the average normal contact force $\langle
  \Fcont \rangle$ (full circles) and on the contact number z (open
  circles). (b) Average critical force $\langle \Fcrit\rangle$ (full
  circles) with respect to the average normal contact force and the penetration
  exponent $\alpha$ (open circles) as functions of the friction coefficient
  $\mu$. The horizontal lines show the reference value $1$ for the exponent
  $\alpha$ (dashed) and for the critical force (solid). The inset shows the
  role of the structural change in the packing. For normally constructed
  packings $\langle \Fcrit\rangle$ is plotted with full squares (new packing is
  constructed for each friction).  Open squares were obtained by changing
  the friction in a fixed packing configuration.}
\label{Fig-ForceAlphaFriction}
\end{figure}

It is important to note that the packings, that are subjected to
perturbations, are newly generated for each value of friction therefore
different values of $\mu$ are accompanied by different packing structures.
In Fig.~\ref{Fig-ForceAlphaFriction}.a the average contact number $z$
\footnote{Rattlers (particles without force-carrying contacts) are
disregarded in $z$.} and the average contact force $\langle \Fcont \rangle$
is plotted. In $\Fcont$ we take only the normal component of the contact
force into account, the average $\langle \, \, \rangle$ is meant over all the
contacts of the given packing. The contact forces are measured in units $F_0$
set by the external pressure and by the average radius of the particles,
$F_0 = 2 R_\text{avg} P_\text{ext}$. The behavior of the packing fraction
(not shown) is similar to that of $z$, its value is 0.84 (0.80) in the
low (high) friction limit. Quantities, that describe the
properties of the packing, have the common behavior that they exhibit
plateaus for low and high friction and show a smooth and monotonic
transition in between.

We find a completely different behavior in the response
(Fig.~\ref{Fig-ForceAlphaFriction}.b).  Both $\alpha$ and the average
critical force $\langle \Fcrit \rangle$ are nonmonotonic functions of the
friction coefficient and exhibit a quite sharp peak at the same place
$\mu\approx 0.1$. This is a remarkable behavior because all systems are
subjected to exactly the same compaction and perturbation procedures,
friction was the only quantity that has been changed.  

When the friction is increased starting from zero, packings are getting
stronger against the perturbation and the induced rearrangements become
more localized. At $\mu \approx 0.1$ the process takes a sharp turn
and further increase of the friction leads to softening and delocalization.

The nonmonotonic behavior and the position of the maxima in
(Fig.~\ref{Fig-ForceAlphaFriction}.b) are reminiscent of the
force-indeterminacy that was reported in \cite{Unger05}. In order to
clarify what role the force-indeterminacy plays in the mechanical
response we perform two independent numerical
measurements in the same test system: on the one hand we investigate the
force-indeterminacy on the other hand we determine the critical forces. Then
the data are compared on the level of single contacts.

\begin{figure}[t]
\centerline{%
\epsfig{figure=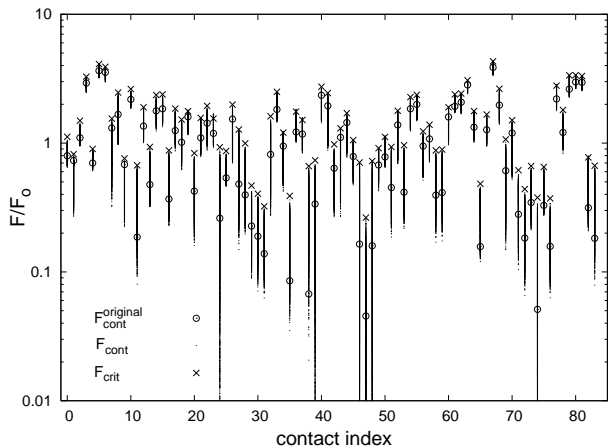,width=0.99\linewidth}}
\caption{The critical force $\Fcrit$ (cross), the original 
contact force (open circle) and the possible values of the normal contact
forces (dots, mostly merged to intervals) are shown for each contact of the
packing.} 
\label{Fig-FcritFcont}
\end{figure}

As mentioned before, in frictional case contact forces are not determined
uniquely but there is an ensemble of force networks that correspond to
equilibrium and to the compatibility conditions (forces are limited by the
Coulomb cone). This ensemble of force networks forms a convex set in the
space of the contact forces \cite{Unger05}. We explore this high
dimensional solution set using a random walk starting with the original
force network. The exploration process is the same as in
\cite{Unger05}. Each step of the random walk provides one possible
force-network. Because the dimension of the solution set is large
(proportional to the number of contacts) only small systems can be handled
efficiently. We generate a packing of $50$ disks with $\mu=0.1$ (total
number of contacts is $85$). As a consequence of the convexity, the
possible values of the normal (the tangential) component of any contact
force are defined by an interval. These intervals are traced out in
Fig.~\ref{Fig-FcritFcont} with help of $20\,000$ force networks that are
collected by the exploration procedure.

The figure shows also the original contact forces (provided by the
compaction process) and the critical forces. The data reveal that the
critical force obtained by the perturbation coincides with the maximum of
the equilibrium solutions at each contact. Thus the critical force can be
directly obtained from the force-ensemble. It also can be seen that a strong
(weak) contact has typically large (small) critical force.

This picture also explains the limit of small and large frictions where the
indeterminacy vanishes. In these cases the length of the force-intervals
goes to zero. Which means that a pair of contacting particles can not
resist a force of separation 
larger than the force itself that originally presses the two contact
surfaces together. This is why the average critical force approaches
$\langle \Fcont \rangle$ on the left and right side of
Fig.~\ref{Fig-ForceAlphaFriction}.b. $\langle \Fcrit
\rangle$ can differ significantly from $\langle \Fcont \rangle$ only if the
indeterminacy of forces is large in the packing.

It was argued in \cite{Unger05} that the nonmonotonic force-indeter\-minacy
can be attributed to two competing effect of the increasing friction:
first, the strengthening of the contacts (increase), and second, the
structural change inside the packing (decrease). Accordingly, if the second
effect is switched off the decay part of the curve $\Fcrit(\mu)$ disappears
as shown by the inset of Fig.~\ref{Fig-ForceAlphaFriction}.b. Here the
response of the same packing configuration was tested for various values of
friction.

In this Letter we reported nontrivial dependence of the mechanical response
on the coefficient of friction and related the results to the indeterminacy
of forces. Nonetheless, many questions arise that urge further studies. We
list three of them here. What is the origin of the displacement exponent
$\alpha$?  Numerical data show strong correlation between $\alpha$ and the
critical force. What is the connection between them?  Finally, how does the
picture, which is presented here for rigid particles, change for deformable
ones?

We acknowledge support by grant OTKA T049403 and \"Oveges project GranKJ06 of
\includegraphics[height=1.5ex]{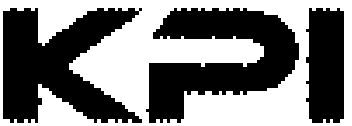} and 
\includegraphics[height=2.5ex]{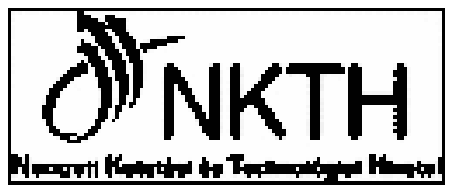}.

\bibliography{Ref}

\end{document}